\documentclass[12pt]{article}
\usepackage{times}
\usepackage{geometry}
\usepackage[utf8]{inputenc}
\usepackage{enumitem,amssymb}
\usepackage{ragged2e}
\newlist{thematic}{itemize}{8}
\setlist[thematic]{label=$\square$}
\usepackage{pifont}

\usepackage{natbib} \bibpunct[]{(}{)}{,}{a}{}{,}
\usepackage{multicol}

\usepackage{xcolor}

\usepackage{graphicx}
\graphicspath{{./}{./figures/}}
\usepackage[breaklinks,colorlinks,allcolors=teal]{hyperref}

\usepackage[footnotesize,bf,FIGBOTCAP,tight]{subfigure}

\newcommand*{\ltsim}{\ {\raise-.75ex\hbox{$\buildrel<\over\sim$}}\ }
\newcommand*{\gtsim}{\ {\raise-.75ex\hbox{$\buildrel>\over\sim$}}\ }
\newcommand*{\proptosim}{\ {\raise-.75ex\hbox{$\buildrel\propto\over\sim$}}\ }

\makeatletter

\renewcommand\section{\@startsection {section}{1}{\z@}%
                                   {-1.0ex \@plus -0.3ex \@minus -0.1ex}%
                                   {0.5ex \@plus.1ex}%
                                   {\normalfont\large\bfseries}}

\makeatother

\begin{document}
\pagestyle{empty}
\raggedright
\Large
Astro2020 Science White Paper \linebreak

The Future Landscape of High-Redshift Galaxy Cluster Science \linebreak
\normalsize

\noindent \textbf{Thematic Areas:}
~$\bullet$ {Cosmology and Fundamental Physics}
~$\bullet$ {Galaxy Evolution}

\bigskip
\justifying 

\noindent
\textbf{Abstract:}

\smallskip

\noindent
Modern galaxy cluster science is a multi-wavelength endeavor with cornerstones provided by X-ray, optical/IR, mm, and radio measurements. In combination, these observations enable the construction of large, clean, complete cluster catalogs, and provide precise redshifts and robust mass calibration. The complementary nature of these multi-wavelength data dramatically reduces the impact of systematic effects that limit the utility of measurements made in any single waveband. The future of multi-wavelength cluster science is compelling, with cluster catalogs set to expand by orders of magnitude in size, and extend, for the first time, into the high-redshift regime where massive, virialized structures first formed. 
Unlocking astrophysical and cosmological insight from the coming catalogs will require new observing facilities that combine high spatial and spectral resolution with large collecting areas, as well as concurrent advances in simulation modeling campaigns.
Together, future multi-wavelength observations will resolve the thermodynamic structure in and around the first groups and clusters, distinguishing the signals from active and star-forming galaxies, and unveiling the interrelated stories of galaxy evolution and structure formation during the epoch of peak cosmic activity.

\raggedright 
\bigskip 

\textbf{Principal Author:}

Name:	Adam B.\ Mantz
 \linebreak						
Institution:  Kavli Institute for Partical Astrophysics and Cosmology, Stanford University
 \linebreak
Email: \href{mailto:amantz@stanford.edu}{\tt amantz@stanford.edu}
 \linebreak
Phone:  +1 650 498 7747
 \linebreak
 
\raggedright 

\textbf{Co-authors:}
  \linebreak
  Steven W. Allen$^{1,2,3}$, 
Nicholas Battaglia$^{4}$, 
Bradford Benson$^{5,6}$, 
Rebecca Canning$^{1,3}$, 
Stefano Ettori$^{7}$, 
August Evrard$^{8}$, 
Anja von der Linden$^{9}$, 
Michael McDonald$^{10}$ 
\\
~\\


\newcommand{\Amherst}{University of Massachusetts, Amherst, MA 01003 USA}
\newcommand{\ANLHEP}{HEP Division, Argonne National Laboratory, Lemont, IL 60439, USA}
\newcommand{\APC}{Laboratoire Astroparticule et Cosmologie (APC), CNRS/IN2P3, Universit\'e Paris Diderot, 10, rue Alice Domon et Léonie Duquet, 75205 Paris Cedex 13, France}
\newcommand{\ASU}{Arizona State University, Tempe, AZ  85287}
\newcommand{\BenGurion}{Department of Physics, Ben-Gurion University, Be'er Sheva 84105, Israel}
\newcommand{\BNL}{Brookhaven National Laboratory, Upton, NY 11973}
\newcommand{\Brown}{Brown University, Providence, RI 02912}
\newcommand{\Bub}{Boston University, Boston, MA 02215}
\newcommand{\BU}{Boston University, Boston, MA 02215}
\newcommand{\Buffalo}{Department of Physics, University at Buffalo, SUNY Buffalo, NY 14260 USA}
\newcommand{\Caltech}{California Institute of Technology, Pasadena, CA 91125}
\newcommand{\Cardiff}{School of Physics and Astronomy, Cardiff University, The Parade, Cardiff, CF24 3AA, UK}
\newcommand{\Carleton}{Carleton University, K1S 5B6 Ottawa, Canada}
\newcommand{\Carnegie}{The Observatories of the Carnegie Institution for Science, 813 Santa Barbara St., Pasadena, CA 91101, USA}
\newcommand{\Cavendish}{Astrophysics Group, Cavendish Laboratory, J.J.Thomson Avenue, Cambridge, CB3 0HE, UK}
\newcommand{\CCA}{Center for Computational Astrophysics, 162 5th Ave, 10010, New York, NY, USA}
\newcommand{\CPPM}{Aix Marseille Univ, CNRS/IN2P3, CPPM, Marseille, France}
\newcommand{\CEADAP}{D\'epartement d’Astrophysique, CEA Saclay DSM/Irfu, 91191 Gif-sur-Yvette, France}
\newcommand{\CERN}{CERN, Geneva, Switzerland}
\newcommand{\CfA}{Harvard-Smithsonian Center for Astrophysics, MA 02138}
\newcommand{\CFT}{Center for Theoretical Physics, Polish Academy of Sciences, al. Lotnik\'{o}w 32/46, 02-668, Warsaw, Poland}
\newcommand{\Cincinnati}{University of Cincinnati, Cincinnati, OH 45221}
\newcommand{\CITA}{Canadian Institute for Theoretical Astrophysics, University of Toronto, Toronto, ON M5S 3H8, Canada}
\newcommand{\CNRSA}{CNRS, Laboratoire d'Annecy-le-Vieux de Physique Th\'{e}orique, France}
\newcommand{\CNYang}{C.N. Yang Institute for Theoretical Physics State University of New York Stony Brook, NY 11794}
\newcommand{\CMUCosmo}{Department 
of Physics, McWilliams Center for Cosmology, Carnegie Mellon University}
\newcommand{\Columbia}{Columbia University, New York, NY 10027}
\newcommand{\Cornell}{Cornell University, Ithaca, NY 14853}
\newcommand{\CPthree}{CP3-Origins, 5230 Odense, Denmark}
\newcommand{\CWRU}{Case Western Reserve University, Cleveland, OH 44106}
\newcommand{\daa}{Department of Astronomy and Astrophysics, University of Toronto, ON, M5S3H4}
\newcommand{\damtp}{DAMTP, Centre for Mathematical Sciences, Wilberforce Road, Cambridge, UK, CB3 0WA}
\newcommand{\DESY}{DESY,  22607 Hamburg, Germany}
\newcommand{\DFI}{Departamento de F\'isica, FCFM, Universidad de Chile, Blanco Encalada 2008, Santiago, Chile}
\newcommand{\DOE}{US. Department of Energy, Germantown, MD 20874}
\newcommand{\drexel}{Drexel University, Philadelphia, PA 19104}
\newcommand{\Duke}{Duke University and Triangle Universitites Nuclear Laboratory, Durham, NC 27708}
\newcommand{\DukePhys}{Department of Physics, Duke University, Durham, NC 27708, USA}
\newcommand{\dunlap}{Dunlap Institute for Astronomy and Astrophysics, University of Toronto, ON, M5S3H4}
\newcommand{\Durham}{Department of Physics, Lower Mountjoy, South Rd, Durham DH1 3LE, United Kingdom}
\newcommand{\ED}{University of Edinburgh, EH8 9YL Edinburgh, United Kingdom}
\newcommand{\EPFL}{Institute of Physics, Laboratory of Astrophysics, Ecole Polytechnique Fédérale de Lausanne (EPFL), Observatoire de Sauverny, 1290 Versoix, Switzerland}
\newcommand{\ETH}{ETH Zurich, Institute for Particle Physics, 8093 Zurich, Switzerland}
\newcommand{\FNAL}{Fermi National Accelerator Laboratory, Batavia, IL 60510}
\newcommand{\FQAUB}{Dept. de F\' isica Qu\` antica i Astrof\' isica, Universitat de Barcelona, Mart\' i i Franqu\` es 1, E08028 Barcelona, Spain}
\newcommand{\FSU}{Florida State University, Tallahassee, FL 32306}
\newcommand{\Glasgow}{University of Glasgow, G12 8QQ Glasgow, United Kingdom}
\newcommand{\GRAPPA}{GRAPPA Institute, University of Amsterdam, Science Park 904, 1098 XH Amsterdam, The Netherlands}
\newcommand{\GSFC}{Goddard Space Flight Center, Greenbelt, MD 20771 USA}
\newcommand{\GWU}{George Washington University, Washington, DC 20052}
\newcommand{\Hampton}{Hampton University, Hampton, VA 23668}
\newcommand{\HarvardPhys}{Department of Physics, Harvard University, Cambridge, MA 02138, USA}
\newcommand{\Haverford}{Haverford College, 370 Lancaster Ave, Haverford PA, 19041, USA}
\newcommand{\Hawaii}{University of Hawaii, Honolulu, HI 96822}
\newcommand{\HKUST}{The Hong Kong University of Science and Technology, Hong Kong SAR, China}
\newcommand{\houston}{University of Houston, Houston, TX 77204}
\newcommand{\IAP}{Institut d'Astrophysique de Paris (IAP), CNRS \& Sorbonne University, Paris, France}
\newcommand{\IAS}{Institute for Advanced Study, Princeton, NJ 08540}
\newcommand{\IBS}{Institute for Basic Science (IBS), Daejeon 34051, Korea}
\newcommand{\ICC}{ICC, University of Barcelona, IEEC-UB, Mart\' i i Franqu\` es, 1, E08028 Barcelona, Spain}
\newcommand{\ICCD}{Institute for Computational Cosmology, Department of Physics, Durham University, South Road, Durham, DH1 3LE, UK}
\newcommand{\ICE}{Institute of Space Sciences (ICE, CSIC), Campus UAB, Carrer de Can Magrans, s/n, 08193 Barcelona, Spain}
\newcommand{\ICRR}{Institute for Cosmic Ray Resaerch, The University of Tokyo, 456 Higashi-Mozumi, Kamioka, Hida, Gifu 506-1205, Japan}
\newcommand{\ICTP}{International Centre for Theoretical Physics, Strada Costiera, 11, I-34151 Trieste, Italy}
\newcommand{\IFAE}{Institut de Fisica d’Altes Energies, The Barcelona Institute of Science and Technology, Campus UAB, 08193 Bellaterra (Barcelona), Spain}
\newcommand{\IFPU}{IFPU - Institute for Fundamental Physics of the Universe, Via Beirut 2, 34014 Trieste, Italy}
\newcommand{\IFT}{Instituto de Fisica Teorica UAM/CSIC, Universidad Autonoma de Madrid, 28049 Madrid, Spain}
\newcommand{\IFUNAM}{IFUNAM - Instituto de F\'{i}sica, Universidad Nacional Aut\'onoma de M\'etico, 04510 CDMX, M\'exico}
\newcommand{\IHEP}{Institute of High Energy Physics, Austrian Academy of Sciences, 1050 Vienna, Austria}
\newcommand{\Imperial}{Theoretical Physics, Blackett Laboratory, Imperial College, London, SW7 2AZ, U.K.}
\newcommand{\Indiana}{Indiana University, Bloomington, IN 47405}
\newcommand{\INAFOATs}{INAF - Osservatorio Astronomico di Trieste, Via G.B. Tiepolo 11, 34143 Trieste, Italy}
\newcommand{\INAFOAS}{INAF - Osservatorio di Astrofisica e Scienza dello Spazio di Bologna, via Piero Gobetti 93/3, I-40129 Bologna, Italy}
\newcommand{\INFNCag}{Istituto Nazionale di Fisica Nucleare, Sezione di Cagliari,  09126 Cagliari, Italy}
\newcommand{\INFNCat}{Istituto Nazionale di Fisica Nucleare, Sezione di Catania, 95125 Catania, Italy}
\newcommand{\INFNG}{Istituto Nazionale di Fisica Nucleare, Sezione di Genova, 16146 Genova, Italy}
\newcommand{\INFN}{INFN – National Institute for Nuclear Physics, Via Valerio 2, I-34127 Trieste, Italy}
\newcommand{\INFNFE}{Istituto Nazionale di Fisica Nucleare, Sezione di Ferrara, 40122, Italy }
\newcommand{\INFNLNF}{Istituto Nazionale di Fisica Nucleare, Laboratori Nazionali di Frascati, 00044 Frascati, Italy}
\newcommand{\INFNLNS}{Istituto Nazionale di Fisica Nucleare, Laboratori Nazionali del Sud, 95125 Catania, Italy}
\newcommand{\INFNN}{Istituto Nazionale di Fisica Nucleare, Sezione di Napoli, 80125 Napoli, Italy }
\newcommand{\INFNRM}{Istituto Nazionale di Fisica Nucleare, Sezione di Roma, 00185 Roma, Italy}
\newcommand{\INFNT}{Istituto Nazionale di Fisica Nucleare, Sezione di Torino, 10125, Italy }
\newcommand{\ioa}{Institute of Astronomy, University of Cambridge, Cambridge CB3 0HA, UK}
\newcommand{\IPP}{Institute for Particle Physics, BC V8W 3P6 Victoria, Canada}
\newcommand{\IPMU}{Kavli Insitute for the Physics and Mathematics of the Universe (WPI), University of Tokyo, 277-8583 Kashiwa , Japan}
\newcommand{\IPNL}{Universit\'e de Lyon, F-69622, Lyon, France; Universit\'e de Lyon 1, Villeurbanne; CNRS/IN2P3, Institut de Physique Nucl\'eaire de Lyon}
\newcommand{\IRFU}{IRFU, CEA, Universit\'e Paris-Saclay, F-91191 Gif-sur-Yvette, France}
\newcommand{\ITFA}{Institute for Theoretical Physics, University of Amsterdam, Science Park 904, 1098 XH Amsterdam, The Netherlands}
\newcommand{\IUCAA}{The Inter-University Centre for Astronomy and Astrophysics, Pune, 411007, India}
\newcommand{\Jerusalem}{Hebrew University of Jerusalem, 91904 Jerusalem, Israel}
\newcommand{\JHU}{Johns Hopkins University, Baltimore, MD 21218}
\newcommand{\JLAB}{Thomas Jefferson National Laboratory, Newport News, VA 23606}
\newcommand{\JPL}{Jet Propulsion Laboratory, California Institute of Technology, Pasadena, CA, USA}
\newcommand{\KASSI}{Korea Astronomy and Space Science Institute, Daejeon 34055, Korea}
\newcommand{\kavli}{Kavli Institute for Cosmology, Cambridge, UK, CB3 0HA}
\newcommand{\KIAS}{School of Physics, Korea Institute for Advanced Study, 85 Hoegiro, Dongdaemun-gu, Seoul 130-722, Korea}
\newcommand{\KICP}{Kavli Institute for Cosmological Physics, Chicago, IL 60637}
\newcommand{\KIPAC}{Kavli Institute for Particle Astrophysics and Cosmology, Stanford 94305}
\newcommand{\KINGS}{King's College London, WC2R 2LS London, United Kingdom}
\newcommand{\Kobe}{Kobe University, 657-8501 Kobe, Japan}
\newcommand{\KPH}{Johannes Gutenberg University, 55128 Mainz, Germany}
\newcommand{\KPMU}{University of Tokyo, 277-8583  Kashiwa , Japan}
\newcommand{\KSU}{Kansas State University, Manhattan, KS 66506}
\newcommand{\Lafayette}{Lafayette College, Easton, PA 18042}
\newcommand{\LANL}{Los Alamos National Laboratory, Los Alamos, NM 87545}
\newcommand{\LBL}{Lawrence Berkeley National Laboratory, Berkeley, CA 94720}
\newcommand{\Leiden}{Lorentz Institute, Leiden University, Niels Bohrweg 2,Leiden, NL 2333 CA, The Netherlands}
\newcommand{\Liverpool}{University of Liverpool,  L69 7ZE Liverpool , United Kingdom}
\newcommand{\LLNL}{Lawrence Livermore National Laboratory, Livermore, CA, 94550}
\newcommand{\LPC}{Universit\'e Clermont Auvergne, CNRS/IN2P3, Laboratoire de Physique de Clermont, F-63000 Clermont-Ferrand, France}
\newcommand{\LPNHE}{Sorbonne Universit\'e, Universit\'e Paris Diderot, CNRS/IN2P3, Laboratoire de Physique Nucl\'eaire et de Hautes Energies, LPNHE, 4 Place Jussieu, F-75252 Paris, France}
\newcommand{\McGill}{McGill University, Montreal, QC H3A 2T8, Canada}
\newcommand{\Melbourne}{School of Physics, The University of Melbourne, Parkville, VIC 3010, Australia}
\newcommand{\Mines}{Colorado School of Mines, Golden, CO 80401}
\newcommand{\MIT}{Massachusetts Institute of Technology, Cambridge, MA 02139}
\newcommand{\MPE}{Max-Planck-Institut f\"{u}r extraterrestrische Physik (MPE), Giessenbachstrasse 1, D-85748 Garching bei M\"unchen, Germany}
\newcommand{\MPIA}{Max-Planck-Institut f\"{u}r Astrophysik, Karl-Schwarzschild-Str. 1, 85741 Garching, Germany}
\newcommand{\MPP}{Max-Planck-Institut f\"{u}r Physik (Werner-Heisenberg-Institut), F\"ohringer Ring 6, D-80805 M\"unchen, Germany}
\newcommand{\LUPM}{Laboratoire Univers et Particules de Montpellier, Univ. Montpellier and CNRS, 34090 Montpellier, France}
\newcommand{\NAOC}{National Astronomical Observatories, Chinese Academy of Sciences, PR China}
\newcommand{\NCBJ}{National Center for Nuclear Research, Ul.Pasteura 7,Warsaw, Poland}
\newcommand{\NCU}{National Central University, Taoyuan City 32001, Taiwan (R.O.C.)}
\newcommand{\NCSU}{Physics Department, North Carolina State Universitym, 2401 Stinson Dr, Raleigh, NC 27607}
\newcommand{\ND}{University of Notre Dame,vNotre Dame, IN 46556}
\newcommand{\NIU}{Northern Illinois University, DeKalb, Illinois 60115}
\newcommand{\NMSU}{New Mexico State University, Las Cruces, NM 88003}
\newcommand{\NOAO}{National Optical Astronomy Observatory, 950 N. Cherry Ave., Tucson, AZ 85719 USA}
\newcommand{\Northwestern}{Northwestern University, Evanston, IL 60201}
\newcommand{\Nottingham}{University of Nottingham, NG7 2RD Nottingham, United Kingdom}
\newcommand{\NWU}{Northwestern University, Evanston, IL 60208}
\newcommand{\NYU}{New York University, New York, NY 10003}
\newcommand{\OK}{ University of Oklahoma, Norman, OK 73019}
\newcommand{\ORNL}{Oak Ridge National Laboratory, Oak Ridge, TN 37831}
\newcommand{\OSU}{The Ohio State University, Columbus, OH 43212}
\newcommand{\OU}{Department of Physics and Astronomy, Ohio University, Clippinger Labs, Athens, OH 45701, USA}
\newcommand{\OskarKlein}{Oskar Klein Centre for Cosmoparticle Physics, Stockholm University, AlbaNova, Stockholm SE-106 91, Sweden}
\newcommand{\Oxford}{The University of Oxford, Oxford OX1 3RH, UK}
\newcommand{\Oxy}{Occidental College, Los Angeles, CA 90041}
\newcommand{\ParisSud}{Universit\'{e} Paris-Sud, LAL, UMR 8607, F-91898 Orsay Cedex, France \& CNRS/IN2P3, F-91405 Orsay, France}
\newcommand{\PI}{Perimeter Institute, Waterloo, Ontario N2L 2Y5, Canada}
\newcommand{\Pitt}{University of Pittsburgh and PITT PACC, Pittsburgh, PA 15260}
\newcommand{\PNNL}{Pacific Northwest National Laboratory ,Richland, WA 99352}
\newcommand{\PNPI}{Petersburg Nuclear Physics Institute, 188300 Gatchina, Russia}
\newcommand{\Port}{Institute of Cosmology \& Gravitation, University of Portsmouth, Dennis Sciama Building, Burnaby Road, Portsmouth PO1 3FX, UK}
\newcommand{\Princeton}{Princeton University, Princeton, NJ 08544}
\newcommand{\PSU}{The Pennsylvania State University, University Park, PA 16802}
\newcommand{\Purdue}{Purdue University, West Lafayette, IN 47907}
\newcommand{\PW}{Participation Worldscope, Sedona, Arizona and Hong Kong, SAR PRC}
\newcommand{\Queens}{Queen's University , K7L 3N6 Kingston, Canada}
\newcommand{\Queensland}{The University of Queensland, School of Mathematics and Physics, QLD 4072, Australia}
\newcommand{\QMUL}{Queen Mary University of London, Mile End Road, London E1 4NS, United Kingdom}
\newcommand{\RAL}{Radio Astronomy Laboratory, University of California Berkeley, Berkeley, CA 94720, USA}
\newcommand{\Rice}{Department of Physics \& Astronomy, Rice University, Houston, Texas 77005, USA}
\newcommand{\RIT}{Rochester Institute of Technology}
\newcommand{\RomaS}{Dipartimento di Fisica, Universit\`{a} La Sapienza, P. le A. Moro 2, Roma, Italy}
\newcommand{\RUG}{Kapteyn Astronomical Institute, University of Groningen, P.O. Box 800, 9700 AV Groningen, The Netherlands}
\newcommand{\Rutgers}{Department of Physics and Astronomy, Rutgers, the State University of New Jersey, 136 Frelinghuysen Road, Piscataway, NJ 08854, USA}
\newcommand{\Sanford}{Sanford Underground Research Facility, Lead, SD 57754}
\newcommand{\Sassari}{Universit\`a di Sassari, 07100 Sassari,  Italy}
\newcommand{\SCIPP}{University of California at Santa Cruz, Santa Cruz, CA 95064}
\newcommand{\Sejong}{Department of Physics and Astronomy, Sejong University, Seoul, 143-747, Korea}
\newcommand{\Sheffield}{University of Sheffield, S3 7RH Sheffield, United Kingdom}
\newcommand{\SHAO}{Shanghai Astronomical Observatory (SHAO), Nandan Road 80, Shanghai 200030, China}
\newcommand{\Siena}{Siena College, 515 Loudon Road, Loudonville, NY 12211, USA}
\newcommand{\SISSA}{SISSA - International School for Advanced Studies, Via Bonomea 265, 34136 Trieste, Italy}
\newcommand{\SimonFraser}{Department of Physics, Simon Fraser University, Burnaby, British Columbia, Canada V5A 1S6}
\newcommand{\SLAC}{SLAC National Accelerator Laboratory, Menlo Park, CA 94025}
\newcommand{\SMU}{Southern Methodist University, Dallas, TX 75275}
\newcommand{\SNOLAB}{SNOLAB, Lively, ON P3Y 1N2, Canada}
\newcommand{\SoCal}{University of Southern California, CA 90089 }
\newcommand{\Stanford}{Stanford University, Stanford, CA 94305}
\newcommand{\StonyBrook}{Stony Brook University, Stony Brook, NY 11794}
\newcommand{\STSCI}{Space Telescope Science Institute, Baltimore, MD 21218}
\newcommand{\SUNYA}{University at Albany SUNY, Albany, NY 12222}
\newcommand{\SussexAstronomy}{Astronomy Centre, School of Mathematical and Physical Sciences, University of Sussex, Brighton BN1 9QH, United Kingdom}
\newcommand{\Syracuse}{Syracuse University, Syracuse, NY 13244}
\newcommand{\Tamu}{Texas AandM University, College Station, TX 77843 }
\newcommand{\Techsource}{Techsource Incorporated, Los Alamos, NM 87544}
\newcommand{\TelAviv}{Tel-Aviv University,  69978 Tel-Aviv, Israel}
\newcommand{\Temple}{Temple University, Philadelphia, PA 19122}
\newcommand{\TIFR}{Tata Institute of Fundamental Research, Homi Bhabha Road, Mumbai 400005 India}
\newcommand{\Tsinghua}{Department of Physics and Tsinghua Center for Astrophysics, Tsinghua University, Beijing 100084, P R China}
\newcommand{\TUM}{Technical University of Munich,  80333 Munich, Germany}
\newcommand{\UA}{University of Alabama, Tuscaloosa, AL 35487}
\newcommand{\UAS}{Department of Astronomy/Steward Observatory, University of Arizona, Tucson, AZ  85721}
\newcommand{\UAM}{Universidad Aut\'onoma de Madrid, 28049, Madrid, Spain}
\newcommand{\UBC}{University of British Columbia, Vancouver, BC V6T 1Z1, Canada}
\newcommand{\UCB}{Department of Astronomy, University of California Berkeley, Berkeley, CA 94720, USA}
\newcommand{\UCBP}{Department of Physics, University of California Berkeley, Berkeley, CA 94720, USA}
\newcommand{\UCBSSL}{Space Sciences Laboratory, University of California Berkeley, Berkeley, CA 94720, USA}
\newcommand{\UCD}{University of California at Davis, Davis, CA 95616}
\newcommand{\UChicago}{University of Chicago, Chicago, IL 60637}
\newcommand{\UCI}{University of California, Irvine, CA 92697}
\newcommand{\UCLA}{University of California at Los Angeles, Los Angeles,  CA 90095}
\newcommand{\UCL}{University College London, WC1E 6BT London, United Kingdom}
\newcommand{\UCR}{University of California at Riverside, Riverside, CA 92521}
\newcommand{\UCSB}{University of California at Santa Barbara, Santa Barbara, CA 93106}
\newcommand{\UCSC}{University of California at Santa Cruz, Santa Cruz, CA 95064}
\newcommand{\UCSD}{University of California San Diego, La Jolla, CA 92093}
\newcommand{\UFL}{University of Florida, Gainesville, FL 32611}
\newcommand{\UFN}{Universit\`a Federico II di Napoli, 80125 Napoli, Italy}
\newcommand{\UGTO}{Divisi\'on de Ciencias e Ingenier\'ias, Universidad de Guanajuato, Le\'on 37150, M\'exico}
\newcommand{\UKY}{University of Kentucky, Lexington, KY 40506}
\newcommand{\UMD}{University of Maryland, College Park, MD 20742}
\newcommand{\UMiami}{University of Miami, Coral Gables, FL 33124}
\newcommand{\UMich}{University of Michigan, Ann Arbor, MI 48109}
\newcommand{\UMN}{University of Minnesota, Minneapolis, MN 55455}
\newcommand{\UnB}{Instituto de F\'{i}sica, Universidade de Bras\'{i}lia, 70919-970, Bras\'{i}lia, DF, Brazil}
\newcommand{\UNC}{University of North Carolina at Chapel Hill, Chapel Hill, NC 27599}
\newcommand{\UNH}{University of New Hampshire, Durham, NH 03824}
\newcommand{\UNIMI}{Dipartimento di Fisica ``Aldo Pontremoli'', Universit\`a{} degli Studi di Milano, via Celoria 16, 20133 Milano, Italy}
\newcommand{\UNIPD}{Dipartimento di Fisica e Astronomia ``G. Galilei'',Universit\`a degli Studi di Padova, via Marzolo 8, I-35131, Padova, Italy}
\newcommand{\UNM}{University of New Mexico, Albuquerque, NM 87131}
\newcommand{\UNV}{University of Nevada, Reno, NV 89557}
\newcommand{\UoM}{Jodrell Bank Center for Astrophysics, School of Physics and Astronomy, University of Manchester, Oxford Road, Manchester, M13 9PL, UK}
\newcommand{\UPenn}{Department of Physics and Astronomy, University of Pennsylvania, Philadelphia, Pennsylvania 19104, USA}
\newcommand{\UR}{Department of Physics and Astronomy, University of Rochester, 500 Joseph C. Wilson Boulevard, Rochester, NY 14627, USA}
\newcommand{\UrbanaC}{Department of Physics, University of Illinois at Urbana-Champaign, Urbana, Illinois 61801, USA}
\newcommand{\USC}{The University of South Carolina, Columbia, SC 29208}
\newcommand{\USD}{The University of South Dakota, Vermillion, SD 57069}
\newcommand{\UTD}{University of Texas at Dallas, Texas 75080}
\newcommand{\Utenn}{The University of Tennessee, Knoxville, TN 37996}
\newcommand{\Utah}{University of Utah, Department of Physics and Astronomy, 115 S 1400 E, Salt Lake City, UT 84112, USA}
\newcommand{\UVA}{University of Virginia, Charlottesville, VA 22903}
\newcommand{\Uvic}{University of Victoria, BC V8P 5C2 Victoria, Canada}
\newcommand{\UWaterloo}{Department of Physics and Astronomy, University of Waterloo, 200 University Ave W, Waterloo, ON N2L 3G1, Canada}
\newcommand{\UWMadison}{Department of Physics, University of Wisconsin - Madison, Madison, WI 53706}
\newcommand{\UW}{University of Washington, Seattle 98195}
\newcommand{\UWC}{Department of Physics \& Astronomy, University of the Western Cape, Cape Town 7535, South Africa}
\newcommand{\Vanderbilt}{Physics \& Astronomy Department, Vanderbilt University, PMB 401807, 2301 Vanderbilt Place, Nashville, TN 37235}
\newcommand{\VSI}{Van Swinderen Institute for Particle Physics and Gravity, University of Groningen, Nijenborgh 4, 9747~AG~Groningen, The~Netherlands}
\newcommand{\VT}{Virginia Tech, Blacksburg, VA 24061}
\newcommand{\VUU}{Virginia Union University, Richmond, Virginia, 23220}
\newcommand{\WCA}{Centre for Astrophysics, University of Waterloo, Waterloo, Ontario N2L 3G1, Canada}
\newcommand{\Weizmann}{Weizmann Institute of Science, 76100 Rehovot, Israel}
\newcommand{\Wellesley}{Wellesley College, Wellesley, MA 02481}
\newcommand{\wiscIce}{University of Wisconsin, Madison, WI 53706}
\newcommand{\WM}{College of William and Mary, Newport News, VA 23606}
\newcommand{\WUSL}{Washington University in St Louis, St. Louis, MO 63130}
\newcommand{\WVU}{CSEE, West Virginia University, Morgantown, WV 26505, USA}
\newcommand{\WVUGWAC}{Center for Gravitational Waves and Cosmology, West Virginia University, Morgantown, WV 26505, USA}
\newcommand{\Wyoming}{Department of Physics and Astronomy, University of Wyoming, Laramie, WY 82071, USA}
\newcommand{\Yale}{Department of Physics, Yale University, New Haven, CT 06520}
\newcommand{\YorkU}{Department of Physics and Astronomy, York University, Toronto, Ontario M3J 1P3, Canada}

\newcommand{\IRAP}{IRAP, Universit\'e de Toulouse, CNRS, CNES, UPS, Toulouse, France}
\newcommand{\AIfA}{Argelander Institute for Astronomy, University of Bonn, Auf dem H\"ugel 71, D-53121 Bonn, Germany}
\newcommand{\Hamburg}{Hamburger Sternwarte, Gojenbergsweg 112, 21029 Hamburg, Germany}
\newcommand{\RIKEN}{Computational Astrophysics Laboratory -- RIKEN, 2-1 Hirosawa, Wako, Saitama 351-0198, Japan}
\newcommand{\intwopthree}{IN2P3 Computing Center, CNRS, Lyon-Villeurbanne, France}
\newcommand{\LeidenObs}{Leiden Observatory, Leiden University, PO Box 9513, 2300 RA Leiden, The Netherlands}
\newcommand{\LMU}{Ludwig-Maximilians-Universit\"at, Scheinerstr. 1, 81679 Munich, Germany}
\newcommand{\Chicago}{The University of Chicago, Chicago, IL 60637}
\newcommand{\Lagrange}{Laboratoire Lagrange, UMR 7293, Universit\'e de Nice Sophia Antipolis, CNRS, Observatoire de la C\^ote d'Azur, 06304 Nice, France}
\newcommand{\MSU}{Michigan State University, East Lansing, MI 48824-2320, USA}
\newcommand{\Bristol}{University of Bristol, Tyndall Ave, Bristol BS8 1TL, UK}
\newcommand{\CEA}{Service d'Astrophysique, CEA Saclay, Orme des Merisiers, F-91191 Gif-sur-Yvette cedex, France}
\newcommand{\Heidelberg}{Astronomisches Rechen-Institut, Zentrum f\"ur Astronomie der Universit\"at Heidelberg, M\"onchhofstrasse 12-14, D-69120 Heidelberg, Germany}
\newcommand{\ESO}{European Southern Observatory, Garching, Germany}
\newcommand{\IFRU}{IRFU, CEA, Universit\'e Paris-Saclay, 91191, Gif-Sur-Yvette, France}
\newcommand{\UND}{University of North Dakota, Grand Forks, ND 58202}
\renewcommand{\CfA}{Harvard-Smithsonian Center for Astrophysics, Cambridge, MA 02138}
\newcommand{\HDSI}{Harvard Data Science Initiative, Harvard University, Cambridge, MA 02138}
\newcommand{\Bologna}{Universit\`a di Bologna, via Gobetti 93/2, 40129 Bologna, Italy}
\newcommand{\Bonn}{University of Bonn, Bonn, Germany}
\newcommand{\Ljubljana}{University of Ljubljana, Jadranska 19, 1000 Ljubljana, Slovenia}
\newcommand{\Boulder}{University of Colorado, Boulder, CO 80309, USA}
\newcommand{\AMES}{NASA Ames Research Center, Moffett Field, CA 94035, USA}
\newcommand{\UCAS}{University of Chinese Academy of Sciences, Beijing 100049, China}
\newcommand{\Alpes}{Univ.\ Grenoble Alpes, Univ. Savoie Mont Blanc, CNRS, LAPP, 74000 Annecy, France}
\newcommand{\Nara}{Department of Physics, Nara Women’s University, Kitauoyanishi-machi, Nara, Nara 630-8506, Japan}
\newcommand{\UCDenver}{University of Colorado, Denver, CO 80204, USA}
\newcommand{\Swinburne}{Swinburne University of Technology, Hawthorn, Victoria 3122, Australia}
\newcommand{\MTAeotvos}{MTA-E\"otv\"os University Lend\"ulet Hot Universe Research Group, P\'azm\'any P\'eter s\'et\'any 1/A, Budapest, 1117, Hungary}
\newcommand{\Masaryk}{Department of Theoretical Physics and Astrophysics, Faculty of Science, Masaryk University, Kotl\'a\v{r}sk\'a 2, Brno, 611 37, Czech Republic}
\newcommand{\Hiroshima}{School of Science, Hiroshima University, 1-3-1 Kagamiyama, Higashi-Hiroshima 739-8526, Japan}
\newcommand{\Dartmouth}{Dartmouth College, Hanover, NH 03755}
\newcommand{\IKI}{Space Research Institute (IKI), Profsoyuznaya 84/32, Moscow 117997,Russia
}
\newcommand{\SRON}{SRON Netherlands Institute for Space Research, Landleven 12, 9747 AD, Groningen, The Netherlands}
\newcommand{\IIT}{IIT Hyderabad, Kandi, Telangana 502285, India}

\textbf{Endorsers:}
  \linebreak
Muntazir Abidi$^{11}$, 
Zeeshan Ahmed$^{2}$, 
Mustafa A. Amin$^{12}$, 
Behzad Ansarinejad$^{13}$, 
Robert Armstrong$^{14}$, 
Camille Avestruz$^{5}$, 
Carlo Baccigalupi$^{15,16,17}$, 
Kevin Bandura$^{18,19}$, 
Wayne Barkhouse$^{20}$, 
Kaustuv moni Basu$^{21}$
% ...
\medskip\\{\small
$^{1}$ \Stanford \\
$^{2}$ \SLAC \\
$^{3}$ \KIPAC \\
$^{4}$ \Cornell \\
$^{5}$ \KICP \\
$^{6}$ \FNAL \\
$^{7}$ \INAFOAS \\
$^{8}$ \UMich \\
$^{9}$ \StonyBrook \\
$^{10}$ \MIT \\
$^{11}$ \damtp \\
$^{12}$ \Rice \\
$^{13}$ \Durham \\
$^{14}$ \LLNL \\
$^{15}$ \SISSA \\
$^{16}$ \IFPU \\
$^{17}$ \INFN \\
$^{18}$ \WVU \\
$^{19}$ \WVUGWAC \\
$^{20}$ \UND \\
$^{21}$ \Bonn \\
% ...
}

\justifying

\pagebreak
\pagestyle{plain}
\setcounter{page}{1}

\section{Introduction}

Observations of clusters of galaxies provide a powerful probe of cosmology and astrophysics \citep*{Voit0410173, Allen1103.4829, Borgani0906.4370}. Statistical measurements of the evolution of the cluster population over time constrain both the growth of cosmic structure and the expansion history of the Universe. Such observations have played a key role in establishing the current ``concordance'' model of cosmology, in which the mass-energy budget of the Universe is dominated by dark matter and dark energy, with the latter being consistent with a cosmological constant (e.g.\ \citealt{White1993Natur.366..429W, Allen0405340, Vikhlinin0812.2720, Mantz0909.3098}). Clusters are also remarkable astrophysical laboratories, providing unique insights into, e.g., the physics of galaxy evolution \citep{von-der-Linden0909.3522} and structure formation \citep{Simionescu1902.00024, Walker1810.00890}, the role of feedback processes \citep{Fabian1204.4114, McNamara1204.0006}, the history of metal enrichment \citep{Mernier1811.01967}, the nature of dark matter (\citealt{Clowe0608407}), and the physics of hot, diffuse, magnetized plasmas \citep{Markevitch0701821, Brunetti1401.7519, van-Weeren1901.04496}. Clusters also serve as natural gravitational telescopes with which to observe the most distant reaches of the Universe (\citealt{Treu1509.00475}). 

The key observations enabling robust population studies of galaxy clusters are: a sky survey on which cluster finding can be systematically performed with a clean selection function (below), accurate redshift estimates, robust absolute mass calibration (typically provided by weak lensing measurements), and targeted follow-up observations (especially at X-ray wavelengths) to provide precise centers and relative masses for the clusters, and measurements of their dynamical states.

\section{Exploiting multi-wavelength synergies in cluster searches}

Galaxy clusters produce observable signals across the electromagnetic spectrum. At X-ray wavelengths, spatially extended bremsstrahlung emission from the hot intracluster medium (ICM) can be clearly identified. In optical and IR data, we can search for overdensities of galaxies, as well as the red colors typical of cluster members. At mm wavelengths, the spectral distortion of the cosmic microwave background (CMB) due to inverse-Compton scattering with the ICM (the Sunyaev-Zel'dovich or SZ effect) provides a nearly redshift-independent way to find clusters.

The primary observation enabling galaxy cluster science is a sky survey on which cluster finding can be systematically performed, ideally over a large sky area and wide range in redshift. While the construction of cluster catalogs in any single waveband can quickly become a frustrating endeavour hampered by systematic limitations, the complementary nature of X-ray, optical and mm-wavelength data provides direct, observational solutions to most issues. X-ray observations, for example, can provide clean, complete catalogs of clusters, as well as multiple low-scatter mass proxies: quantities that are relatively immune to projection effects, correlating tightly with the true, three-dimensional halo mass. The primary disadvantages of X-ray measurements are the need to make them from space (which brings associated cost and risk), the impact of surface brightness dimming (though this is mild at $z>1$; \citealt{Churazov1502.03269}), and the inability to provide precise absolute mass calibration directly. SZ surveys provide a more uniform selection in redshift, with only their sensitivity determining the mass down to which clusters can in principle be detected. This technique provides our best route for finding clusters at high redshifts, although care is needed to understand the impact on selection of radio- and infrared-emissive cluster galaxies, especially at higher redshifts. Future SZ surveys will also have the ability to provide absolute cluster mass calibration through CMB-cluster gravitational lensing (e.g.\ \citealt{Hu0701276}). Like X-ray surveys, optical and near-infrared (OIR) surveys are most effective at low-to-intermediate redshifts, but have the benefit of finding larger numbers of clusters down to lower masses. 
The primary challenges for optical cluster selection are projection effects (which can lead to overestimated richnesses for some clusters)
and the relatively complex nature of the intrinsic mass-–observable scaling relations.
Nonetheless, optical surveys provide an essential complement to X-ray and SZ data in cluster identification, and uniquely provide essential redshift information (from precise multiband photometry or spectroscopy) and precise absolute mass calibration (through galaxy-cluster lensing). Supporting these observational cornerstones, numerical simulations have emerged as a powerful, complementary tool, providing informative priors on the expected correlations between the measured signals \citep{Stanek0910.1599, Farahi1711.04922, Truong1607.00019}. 

\begin{figure*}
  \centering
  \subfigure{\includegraphics[width=3.1in]{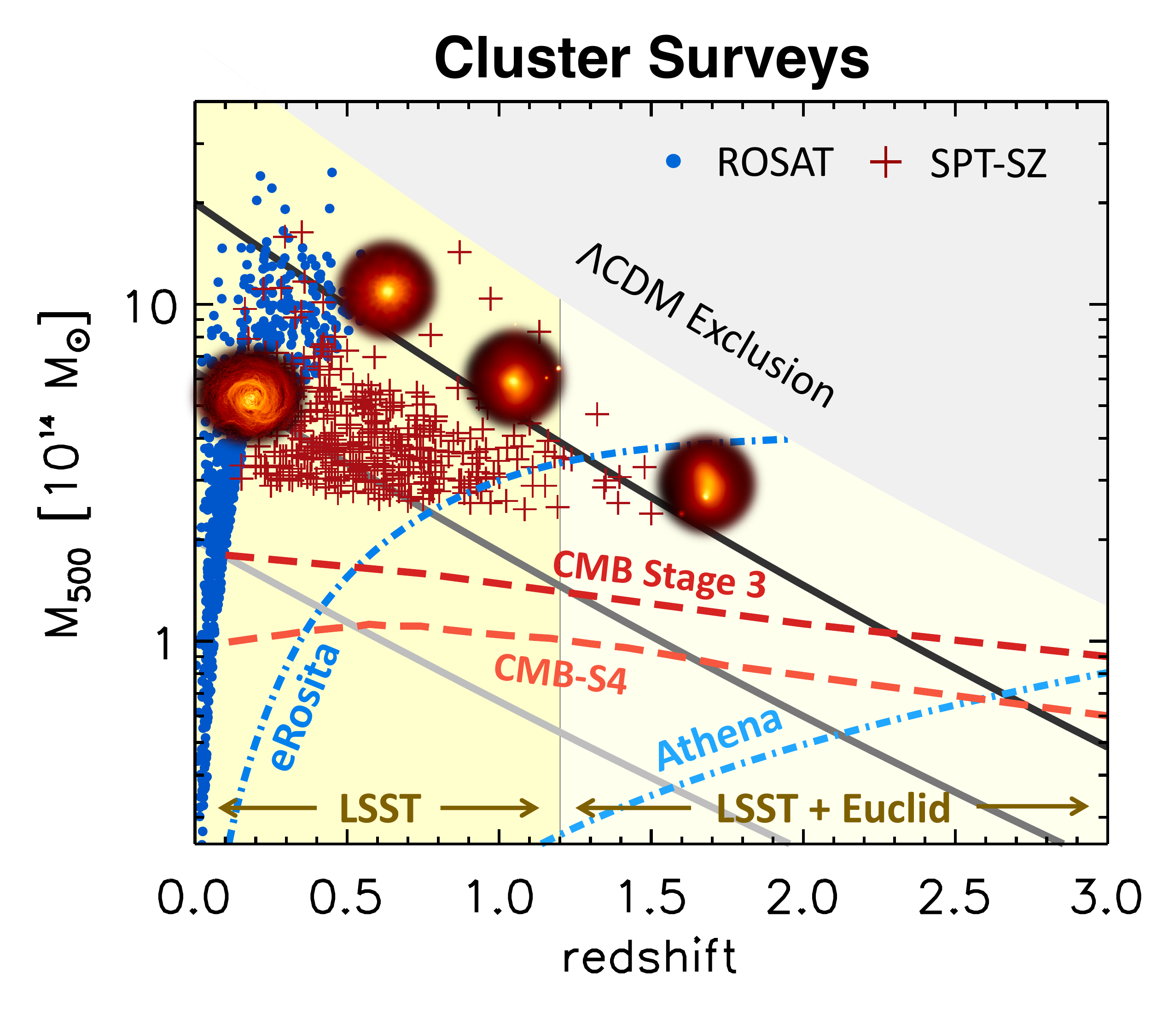}\label{fig:surveys}}
  \hspace{5mm}
  \subfigure{\includegraphics[width=3.05in]{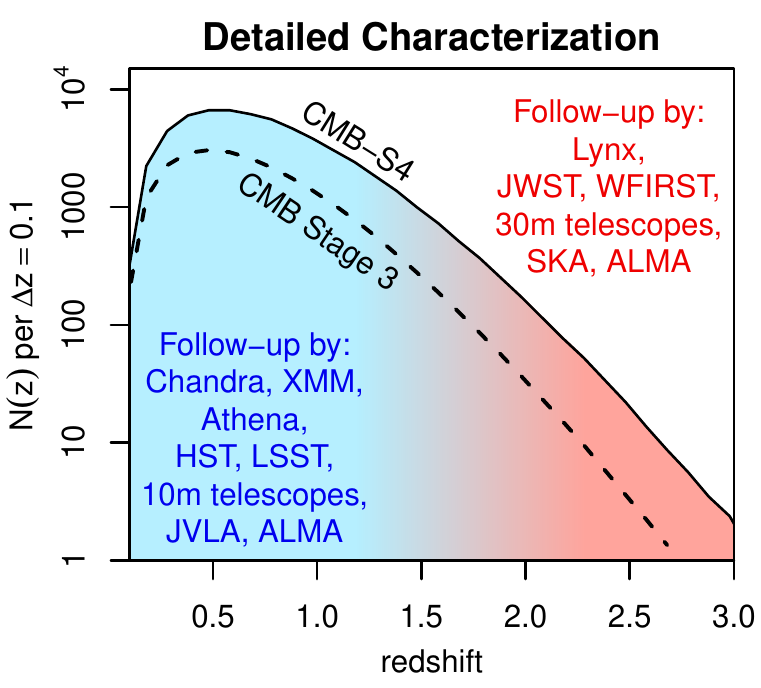}\label{fig:Nz}}
  \vspace{-5mm}
  \caption{\small
    Left: Mass--redshift plot showing some existing cluster catalogs used extensively for astrophysics and cosmology (ROSAT in X-rays, SPT-SZ at mm wavelengths; \citealt{Ebeling0003191, Ebeling1004.4683, Bohringer0703553, Bleem1409.0850}), and the discovery space for the Stage 3 CMB (SPT-3G, Advanced ACT, Simons Observatory), CMB-S4, eROSITA, LSST and {\it Athena} projects. In the standard cosmological model, clusters are not expected to exist in the gray ``exclusion'' region. Solid lines show ``evolutionary'' tracks, tracing out the progenitors of present-day massive clusters.
    Right: The number of SZ cluster detections expected as a function of redshift from Stage 3 SZ surveys and the proposed CMB-S4 project. 
    Blue to red shading shows the transition to the $z\gtsim2$ regime that will be unveiled by new cluster surveys, for which high spatial resolution and throughput are key requirements for extracting information about halo centers, relative masses, dynamical states, internal structure, and galaxy/AGN populations. The proposed new programs will enable the first detailed studies of virialized structure at these redshifts.
  } \label{fig:clusterfinding}
  \vspace{-3mm}
\end{figure*}

Figure~\ref{fig:surveys} illustrates the mass-redshift coverage for two of the leading, current cluster surveys, which have been used extensively for both cosmology and astrophysics studies, and the expected reach of a number of projects, most of which are approved and funded (for more detail see Section~\ref{section:approved}). The figure demonstrates how the forthcoming surveys will vastly increase the size and redshift reach of cluster catalogs, extending out to the epoch when massive clusters first formed and when star formation and AGN activity within them peaked.

Uncovering this distant cluster population is non-trivial. At X-ray wavelengths, it requires an imaging facility with a large collecting area (especially at soft X-ray energies, $<1$\,keV) and sufficient spatial resolution to distinguish truly extended emission from the intracluster medium (ICM) from associations of point-like AGN sources. SZ surveys likewise require a combination of sensitivity and spatial resolution to detect clusters, as well as sufficient frequency coverage to spectrally distinguish measurements of the SZ effect from emissive radio and infrared sources (which contaminate the SZ signal at lower and higher frequencies, respectively). To provide both good redshift estimates and accurate shape measurements for a robust weak lensing mass calibration, optical surveys require exquisite photometric calibration and image quality. To extend the reach of optical measurements significantly beyond $z\gtsim1$, space-based near IR measurements are needed, with sufficient resolution and depth to appropriately complement the optical data.

\begin{figure*}
  \centering
  \includegraphics[width=2.125in]{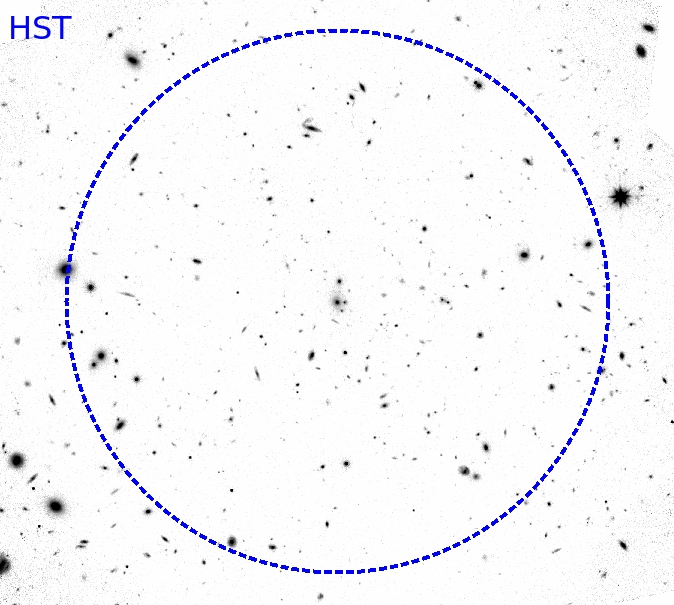}
  \includegraphics[width=2.125in]{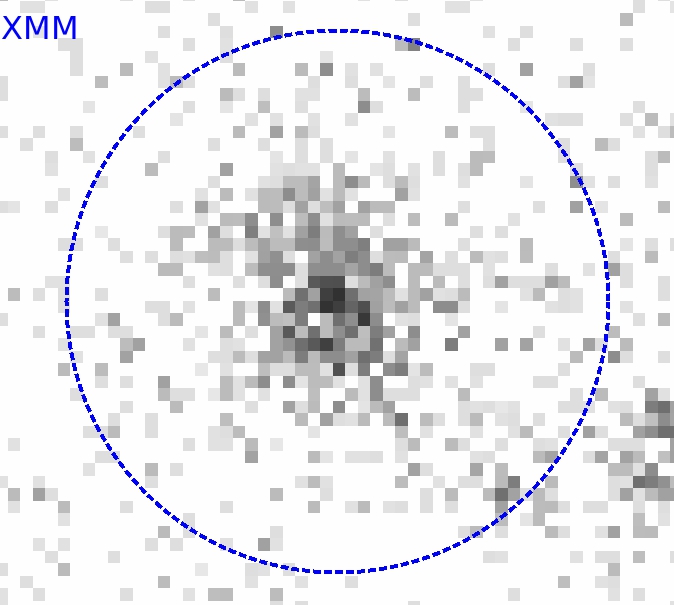}
  \includegraphics[width=2.125in]{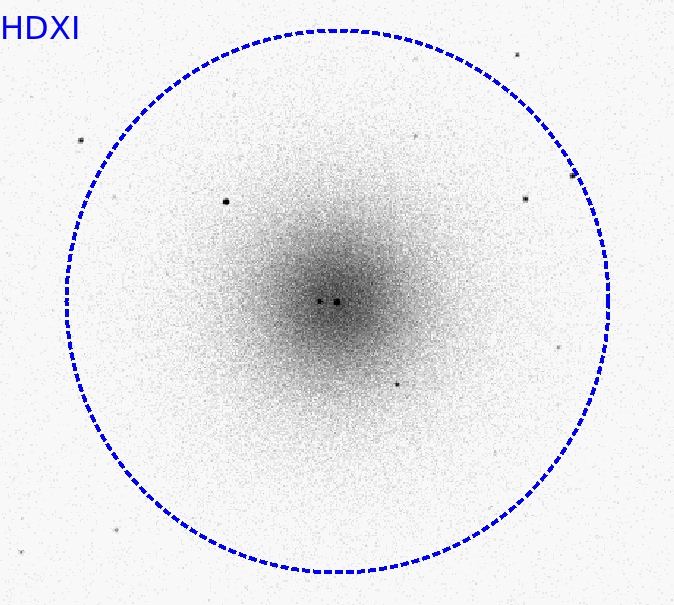}
  \vspace{-3mm}
  \caption{\small
    Images of the $z=2$ cluster XLSSC 122: {\it Hubble} F140W (left; \citealt{Willis-inprep}, in prep), XMM-Newton X-ray (center, 100\,ks), and simulated {\it Lynx} HDXI (right, 100\,ks).
    Dashed circles show the characteristic radius, $r_{200}\sim54''$.
    Realistic densities and luminosities have been generated for cluster and background AGN in the {\it Lynx} simulation, which includes a simple $\beta$ model for the ICM, based on the XMM data.
    Groundbreaking studies of this high-$z$ cluster have benefited from investments of time with XMM, HST, {\it Spitzer}, ALMA, CARMA, and other ground-based observatories (\citealt{Willis1212.4185}; \citealt{Mantz1401.2087, Mantz1703.08221}).
    Such multi-wavelength studies will be routinely superseded by observations with future facilities such as {\it Athena}, JWST, single-dish bolometric mm-wavelength observatories, and 30\,m-class telescopes. High spatial resolution across the electromagnetic spectrum is particularly important for unambiguously identifying galaxy and AGN counterparts.
  } \label{fig:multiw}
  \vspace{-3mm}
\end{figure*}

\section{The Landscape of Approved Projects}
\label{section:approved}

A number of facilities that are approved and in construction will contribute substantially to the future of cluster science. Of special note are the new, dedicated survey instruments: eROSITA in X-rays, LSST and {\it Euclid} at OIR wavelengths, and several ``Stage 3'' ground-based mm-wavelength observatories. Also of note is the {\it Athena} observatory, which will devote a significant fraction of its observing time to performing a deep X-ray survey of several hundred square degrees.

The German-Russian SRG mission, bearing the eROSITA X-ray survey instrument \citep{Merloni1209.3114}, will launch later in 2019. eROSITA will have 30--50 times the sensitivity of the previous all-sky X-ray survey by ROSAT.
Figure~\ref{fig:surveys} shows that the all-sky eROSITA survey is expected to identify essentially all groups out to $z\sim0.3$, all intermediate mass clusters to $\sim0.6$, and the most massive clusters at $z\ltsim2$. 
The FoV-averaged spatial resolution of $26''$, while an improvement over ROSAT, will be limiting at high redshifts, where the angular extent of clusters is small. 
Distinguishing ICM and AGN contributions to the emission from faint, modestly extended sources will require follow-up measurements with higher-spatial-resolution X-ray observatories.

LSST will survey the entire southern sky in {\it ugrizy} over a 10 year period, beginning in 2022. 
It will identify clusters down to the group scale, constrain their redshifts photometrically, and provide precise, stacked weak lensing mass measurements out to a redshift of $\sim1.2$ 
(\citealt{LSSTDESC1211.0310}). Note that the redshift limit reflects the redshift at which the 4000\,\AA\ break moves out of the reddest band. Combining LSST data with near-IR data from {\it Euclid}, an ESA M-class mission scheduled for launch in 2021, will extend the range further. 
Conversely, while {\it Euclid} will identify overdensities of IR-luminous galaxies out to high redshifts \citep{Laureijs1110.3193}, its ability to characterize the cluster population will be enhanced greatly through combination with precise LSST photometry (as well as complementary X-ray and mm observations).  

The ``Stage 3'' CMB (i.e.\ mm-wavelength) surveys most relevant to cluster science are those by SPT-3G, AdvancedACT (both ongoing), and the planned Simons Observatory and CCAT-prime. Taking advantage of the SZ effect, these surveys will break new ground in providing the first large, robustly selected catalogs of clusters at $z>1.5$,  as well as the first informative absolute mass calibration from CMB-cluster lensing. They will find $>3000$ clusters at $z>1$ and $\sim50$ at $z>2$ \citep{Benson1407.2973, De-Bernardis1607.02120, SimonsObs1808.07445, Stacey1807.04354}. However, few detections are expected above $z\sim2.3$ (Fig.\ \ref{fig:clusterfinding}).

{\it Athena}, an ESA mission with NASA involvement, will be the next flagship-class X-ray facility \citep{Nandra1306.2307}. Scheduled for launch in 2031, 
{\it Athena} will combine an order of magnitude increase in effective area compared to XMM-Newton, with a smaller $5''$ (HPD) PSF on axis, degrading only to $\sim10''$ at $30'$ radius. 
{\it Athena}'s grasp significantly exceeds that of any previous X-ray instrument, including eROSITA.
{\it Athena} will also carry the first large, high-spectral-resolution IFU X-ray calorimeter.
With all these advances, we expect to find (\citealt{Zhang-inprep}, in prep) and study \citep{Ettori1306.2322, Pointecouteau1306.2319} very distant galaxy groups and clusters at $z\gtsim2$ over a modest fraction of the sky with {\it Athena}, revolutionizing studies of cluster evolution, dynamics, thermodynamics and metal enrichment. However, due to the small size of these objects (typically $\ltsim50''$ in diameter), these studies will rely on spectral modeling to distinguish emission from AGN and the ICM, rather than directly resolving AGN and small-scale structure within clusters.

\section{New Opportunities}

While the projects described above will undoubtedly transform cluster studies, they are limited in their ability to probe the highest redshifts of interest ($z>2$; due to limited sensitivity and/or sky coverage) and, especially, in their ability to study the astrophysical processes within and around these systems. To do so will require new multi-wavelength facilities with improved sensitivity and enhanced spatial and spectral resolution (Figure~\ref{fig:multiw}).

At X-ray wavelengths, the primary requirement is for an observatory with comparable throughput and spectral capabilities to {\it Athena}, but an order of magnitude higher spatial resolution ($\sim0.5''$). This would open the door to groundbreaking astrophysical measurements, especially (though not exclusively) in the high-$z$ regime (Figure~\ref{fig:Nz}). Recent advances in lightweight, high-resolution, high-throughput X-ray optics have made this goal achievable, as is discussed by the {\it Lynx} and AXIS teams \citep{Zhang2018SPIE10699E..0OZ, Lynx1809.09642}. The ability to spatially resolve and separate AGN within clusters, and to cross-match these sources with ground- and space-based observations in other wavebands, will transform our ability to study how the triggering and quenching of star formation and AGN activity correlates with the evolution of galaxies and their surrounding large scale structure. Resolving the thermodynamic structure and turbulent gas motions within halos, and the distribution of metals within the diffuse cluster gas, will reveal the interwoven stories of galaxy evolution and structure formation, and the roles of feedback from AGN and stars (e.g.\ \citealt{Gaspari1110.6063, McDonald1803.04972}), spanning the epochs when the massive virialized structures first formed and AGN and star formation activity within them peaked. 

For surveys at mm wavelengths, the primary requirements are for greater sensitivity and improved spectral coverage. At high redshifts, even the largest clusters formed have modest spatial extent, making sensitivity and sufficient ($\sim 1'$) spatial resolution the keys to identifying them through the SZ effect, and to providing precise mass calibration from CMB cluster lensing. Adequate spectral coverage is also crucial to separate the SZ effect from emission due to star formation and AGN activity in cluster member galaxies, which are expected to become increasingly important at high redshifts. Configurations such as those being studied for CMB-S4, using multiple, large-aperture telescopes and large, multichroic detector arrays, appear highly promising \citep{CMBS4-Science-Book, Madhavacheril1708.07502}. These measurements would also provide precise (percent-level) absolute mass calibration and similarly precise measurements of the mean pressure and density profiles of the hot gas around clusters (out to many virial radii), from the stacked thermal- and kinetic-SZ signals. 
Follow-up SZ measurements with even higher spatial resolution ($\ltsim10''$) and/or greater spectral coverage (extending above the SZ null) will be possible with ALMA interferometry or single-dish observatories (using successors to the MUSTANG-2 and NIKA-2 instruments and/or new proposed facilities such as CCAT-prime or AtLAST; \citealt{Stacey1807.04354, Mroczkowski1811.02310}).
From space, a new survey such as the proposed PICO mission could build on the legacy of WMAP and Planck, providing all-sky coverage from 20–800\,GHz (albeit with lower spatial resolution than ground-based telescopes), and producing its own catalog of clusters and protoclusters \citep{Hanany1902.10541}.
All these measurements could be complemented by high-spectral resolution X-ray grating spectroscopy of background AGN. Together, these new X-ray and SZ facilities would provide an unprecedented view of the hot, high-redshift Universe.

At OIR wavelengths, WFIRST will provide exquisite data for measuring redshifts and weak lensing of high-$z$ clusters (e.g.\ \citealt{Akeson1902.05569}). These capabilities, along with those of LSST and {\it Euclid}, should be complemented by high-throughput spectrographs with high-multiplexing capabilities on scales of $\sim 10'$.  Such instruments would enable detailed studies of the star-formation and AGN properties of cluster galaxies, spanning the period when they transition from being dominated by star-forming systems to being red-sequence-dominated.  Comprehensive multi-object spectroscopy will also provide a valuable complement to X-ray measurements for dynamical studies of clusters, and will be vital for calibration of photometric redshifts in cluster fields.  

Powerful synergies will also be found at radio wavelengths, where SKA and its precursors (e.g.\ JVLA, LOFAR, MWA, HERA), working in concert with X-ray facilities, will extend studies of AGN feedback out to the highest redshifts. The detection of radio halos and relics, and the correlation of these signals with the dynamic and thermodynamic structure observed at X-ray, optical and mm wavelengths, will reveal the acceleration of particles during subcluster merger events and provide further insight into the virialization process.  

ALMA follow-up will open the door to measurements of molecular gas in high-redshift clusters. At the highest redshifts ($z>4$), 
observations of dusty, star forming galaxies detected by mm surveys  will extend studies of dense environments into the pre-virialized, protocluster regime (e.g.\ \citealt{Miller1804.09231}). Finally, combining the most powerful facilities across all wavelengths, we will continue to use clusters as gravitational cosmic telescopes, to probe the earliest phases of galaxy evolution, and the roles of young stars and AGN in the reionization of the Universe.

Extracting science from more sensitive measurements requires concurrent advances in modeling, including simulations designed to map physical models directly to the space of observable features. Empowering the interpretation of new observational capabilities over the coming decade will require large simulated ensembles of massive halos from cosmological volumes, as well as improvements in resolution and new physical treatments.

\pagebreak
\pagestyle{empty}
\def \mnras {MNRAS}
\def \araa {ARA\&A}
\def \apj {ApJ}
\def \apjs {ApJS}
\def \aap {A\&A} 
\def \asl {Adv.\ Sci.\ Lett.} 
\def \apjl {ApJ} 
\def \prd {Phys.\ Rev.\ D}
\def \physrep {Phys.\ Rep.} 
\def \ssr {Space Sci.\ Rev.}
\def \nat {Nat}

\end{document}